\documentclass[english,reprint,aps,pra, superscriptaddress]{revtex4-1}
\usepackage[T1]{fontenc}
\usepackage[latin9]{inputenc}
\usepackage{amsmath}
\usepackage{amssymb}
\usepackage{graphicx}
\usepackage{braket}
\usepackage[final]{changes}

\makeatletter
\usepackage{babel}

\makeatother

\usepackage{babel}
\usepackage{tikz}
\usetikzlibrary{shapes,arrows}
\usetikzlibrary{arrows.meta}

\tikzstyle{decision} = [diamond, draw, fill=blue!20, 
text width=4.5em, text badly centered, node distance=3cm, inner sep=0pt]
\tikzstyle{block} = [rectangle, draw, fill=blue!30, 
text width=18.5em, text centered, rounded corners, minimum height=4em]
\tikzstyle{line} = [draw, -{Latex[length=0.3cm,width=0.5cm]}, ultra thick]
\tikzstyle{cloud} = [fill=red!50,draw, ellipse, node distance=3cm,
minimum height=2em]

\begin{document}

\title{Free to Harmonic Unitary Transformations in Quantum and Koopman Dynamics}
\date{\today}

\author{Gerard McCaul}
\email{gmccaul@tulane.edu}
\affiliation{Tulane University, New Orleans, LA 70118, USA}

\author{Denys I. Bondar}
\affiliation{Tulane University, New Orleans, LA 70118, USA}
\begin{abstract}
    \added{In the context of quantum dynamics} there exists a coordinate transformation which  maps the free particle to the  harmonic oscillator. Here we extend this result by reformulating it as a unitary operation followed by a time coordinate transformation. We demonstrate that an equivalent transformation can be performed for classical systems in the context of Koopman-von Neumann (KvN) dynamics. We further extend this mapping both to dissipative evolutions as well as for a quantum-classical hybrid, and show that this mapping imparts an identical time-dependent scaling on the dissipation parameters for both types of dynamics. The derived classical procedure presents a number of opportunities to import squeezing dependent quantum procedures (such as Hamiltonian amplification) into the classical regime.     
\end{abstract}
\maketitle

\section{Introduction}
It is a little known fact that there exists an mapping between a quantum free particle and a harmonic oscillator. This formal connection between two distinct systems may seem surprising initially, particularly given that a free particle has a continuous spectrum, while the oscillator's is discrete. Nevertheless, there exists a simple set of coordinate transforms which will transform the wavefunction which solves for the free particle into one which is a solution for a harmonic oscillator \cite{Solov'ev,osti_4333617}. This mapping allows one to transition between free and bound representations of dynamics \cite{qoptics}, which can be useful for obtaining otherwise inaccessible analytic results \cite{PhysRevLett.110.030401}. Moreover, being able to map a free particle to a trapping potential avoids the unbounded spreading that can make numerical simulations of free systems challenging. 

The origin of this mapping lies in quantum optics \cite{Takagi,qoptics}, but has also found use in the analysis of freely moving waves \cite{PhysRevA.89.014103}, inverted oscillators \cite{BARTON1986322,Yuce_2006}, instantaneous transitions \cite{Steuernagel2014}, as well as to the equivalence principle \cite{2207.05073,DHASMANA2021168623}. The mapping itself can be compactly stated - if one has a wave function $\phi(x,t)$ which obeys (setting the mass $m = \hbar$=1 for convenience)
\begin{equation}
    i\frac{\partial \phi(x,t)}{\partial t}=-\frac{1}{2}\frac{\partial^2\phi(x,t)}{\partial x^2},
\end{equation}
this may be brought to the form of a harmonic oscillator using an invertible coordinate transformation given by:
\begin{equation}
    t(\tau)=\tan(\omega \tau),\ \ \ \ \ \ \ x(\xi,\tau)=\frac{\sqrt{\omega}}{\cos(\omega \tau)}\xi.
\end{equation}
This then leads to a harmonic type Sch{\"o}dinger equation:
\begin{equation}
    i\frac{\partial \psi(\xi,\tau)}{\partial \tau}=-\frac{1}{2}\frac{\partial^2\psi(\xi,\tau)}{\partial \xi^2}+\frac{1}{2}\omega^2\xi^2,
\end{equation}
where 
\begin{equation}
\label{eq:wavefunction}
    \psi(\xi,\tau)=\frac{{\rm e}^{\frac{itx^2}{2(t^2+1)}}}{(1+t^2)^{\frac{1}{4}}}\phi(x(\xi,\tau),t(\tau)).
\end{equation}

This mapping may be checked by direct substitution \cite{Steuernagel2014}, and can be extended to both an arbitrary number of spatial dimensions and a time dependent frequency $\omega\to\omega(\tau)$ \cite{Takagi}. 

Significantly, this mapping is quite distinct from either the manner in which quantum harmonic Hamiltonians are diagonalised, or the canonical action-angle transformations which  convert free to harmonic systems in classical dynamics \cite{PhysRevE.99.062121}. The existence of a quantum coordinate transformation of this type naturally raises the question of whether an equivalent exists in the classical case. To investigate this question, we first ask whether the coordinate transform for the Schr{\"o}dinger equation mapping can be performed in a representation-free manner - i.e. whether there exists a unitary operator which is equivalent to the coordinate mapping shown above. In Sec.\ref{Sec:quantum}, we derive the free to harmonic mapping as consisting of a unitary transformation followed by a time coordinate transform. An equivalent procedure is found for classical systems in Sec.\ref{sec:Classical}, where an analogous unitary transforming the Koopman operator is constructed. We further investigate the effect of this mapping on \added{both dissipative and quantum-classical hybrid evolutions} in Sec.\ref{sec:open} and close with a discussion of the results in Sec.\ref{sec:Discussion}.  

\section{Quantum Case \label{Sec:quantum}}
We wish to first investigate whether the coordinate map between a free particle and the harmonic oscillator can be expressed as a unitary transformation, as the existence of such a unitary should provide a transparent method to replicate the free to harmonic map in a classical context. To find this unitary, it is worth considering first both how the coordinates and the wavefunction are transformed.  $x\to\xi$ is a time dependent scaling, while the $\phi\to\psi$ map is strongly reminiscent of a gauge transformation. This immediately suggests that any overall unitary $\hat{\mathcal{U}}$ mapping the free to harmonic system will consist of the composition of two unitary operators.

We begin by considering a scaling unitary $\hat{U}(t)$ of the form: 
\begin{align}
    \hat{U}(t)&={\rm e}^{-i\frac{\ln(c(t))}{2}\left(\hat{x}\hat{p}+\hat{p}\hat{x}\right)}, \\
     \hat{U}^\dagger(t)\hat{x}\hat{U}(t)&=c(t)\hat{x},\\
     \hat{U}^\dagger(t)\hat{p}\hat{U}(t)&=c^{-1}(t)\hat{p}, \\
    i \hat{U}^\dagger(t)\frac{\partial \hat{U}(t)}{\partial t}&=\frac{\dot{c}(t)}{2c(t)}\left(\hat{x}\hat{p}+\hat{p}\hat{x}\right).
\end{align}
This is nothing more than the \textit{squeezing operator}, common in quantum optics but here represented in first quantisation \cite{Schleich2001-ch}. As a time dependent unitary transformation it will transform the system Hamiltonian as
\begin{align}
\hat{H}^\prime= \hat{U}^\dagger(t)\hat{H} \hat{U}(t)-i \hat{U}^\dagger(t)\frac{\partial \hat{U}(t)}{\partial t}.
\end{align}
If we initially have a free-particle system where $\hat{H}=\frac{1}{2}\hat{p}^2$, then the transformed Hamiltonian \added{$H^\prime$} will be given by
\begin{align}
\hat{H}^\prime&= \frac{1}{2c^2(t)}\hat{p}^2-\frac{\dot{c}(t)}{2c(t)}\left(\hat{x}\hat{p}+\hat{p}\hat{x}\right) \notag \\
&= \frac{1}{2c^2(t)}\left(\hat{p}-c(t)\dot{c}(t)\hat{x}\right)^2-\frac{\dot{c}^2(t)}{2}\hat{x}^2. \label{eq:Htransformed}
\end{align}

This scaling unitary has introduced a harmonic potential term, at the cost of a longitudinal vector potential. This can be removed via a unitary gauge transformation $\hat{R}(t)$:
\begin{align}
   \hat{R}(t)&={\rm e}^{i\frac{1}{2}c(t)\dot{c}(t)\hat{x}^2}, \\
     \hat{R}^\dagger(t)\hat{p}\hat{R}(t)&=\hat{p}+c(t)\dot{c}(t)\hat{x}, \\
     -i \hat{R}^\dagger(t)\frac{\partial \hat{R}(t)}{\partial t}& = \frac{1}{2}(\dot{c}^2(t)+c(t)\ddot{c}(t))\hat{x}^2.
\end{align}
Applying this to $\hat{H}^\prime$, we obtain the Hamiltonian:
\begin{equation}
    \hat{H}^{\prime \prime} = \frac{1}{c^2(t)}\left(\frac{1}{2}\hat{p}^2+\frac{1}{2}\ddot{c}(t)c^3(t)\hat{x}^2\right).
\end{equation}

It is interesting to note that while $\hat{H}^{\prime \prime}$  is constructed from unitaries $\hat{R}(t)$ and $\hat{U}(t)$ which clearly do not commute, it is in fact possible to derive an expression for  $\hat{\mathcal{U}}=\hat{U}(t)\hat{R}(t)$ as a single unitary. This is principally due to the fact that we have the commutation relation:
\begin{equation}
    \left[\hat{x}\hat{p}+\hat{p}\hat{x},\hat{x}^2\right]=-4i\hat{x}^2.
\end{equation}
This is relevant as the operator $\hat{Z}$ in ${\rm e}^{\hat{Z}}={\rm e}^{\hat{X}} {\rm e}^{\hat{Y}}$ has a closed form when $[\hat{X},\hat{Y}]=u\hat{Y}$ \cite{VanBrunt2015}. Explicitly this is given by 
\begin{equation}
    \hat{Z}=\hat{X}+f(u)\hat{Y} \label{eq:BCH}
\end{equation}
where
\begin{equation}
f(u) = \frac{u}{2}(1+\coth(\frac{u}{2})).
\end{equation}
Taking $\hat{X}=-i\frac{\ln(c(t))}{2}\left(\hat{x}\hat{p}+\hat{p}\hat{x}\right)$ and $\hat{Y}=i\frac{1}{2}c(t)\dot{c}(t)\hat{x}^2$ results in
\begin{equation}
    [\hat{X},\hat{Y}]=-2\ln(c(t))\hat{Y}.
\end{equation}
From which we obtain
\begin{equation}
  \hat{\mathcal{U}}={\rm e}^{-i\frac{\ln(c(t))}{2}\left(c(t)\dot{c}(t)\left(1-\frac{c^2(t)+1}{c^2(t)-1}\right)\hat{x}^2+\hat{x}\hat{p}+\hat{p}\hat{x}\right)}.
\end{equation}

To make contact with the harmonic oscillator, it is necessary to specify the form of $c(t)$ such that it satisfies $\ddot{c}(t)=c^{-3}(t)\omega^2$. Solving this differential equation gives an expression for $c(t)$:
\begin{equation}
    c(t)=\sqrt{1+\omega^2t^2}.
\end{equation}
The final step is also to transform the time coordinate, using:
\begin{align}
    \tau &= \int^t_0 {\rm d}t^\prime \ \frac{1}{1+\omega^2t^2}=\frac{1}{\omega}\arctan(\omega t), \label{eq:ttransform}\\
    c(\tau) &=\frac{1}{\cos(\omega\tau)}, \label{eq:scaletransform}
\end{align}
such that ${\rm d}t = c^2(t){\rm d} \tau$. This then completes the mapping, which can be summarised as: 
\begin{align}
    i\frac{\partial \ket{\psi}}{\partial t}&=\frac{1}{2}\hat{p}^2\ket{\psi} \\
    \implies i\frac{\partial \ket{\bar{\psi}}}{\partial \tau}&=\frac{1}{2}\left(\hat{p}^2+\omega^2\hat{x}^2\right)\ket{\bar{\psi}} \\
\end{align}
provided that
\begin{equation}
    \omega t=\tan(\omega \tau), \ \ \ \ \ket{\bar{\psi}}=\hat{R}^\dagger(t)\hat{U}^\dagger(t)\ket{\psi}.
\end{equation}

Critically, we find that this mapping has the form of a unitary transformation \textit{in addition} to a time coordinate transformation. As has been noted in Ref. \cite{Steuernagel2014}, this latter transform plays a particularly important role as the discontinuities of the $\tan$ function are what effect the transition from the continuous free particle energetic spectrum to the discrete spectrum of the oscillator. This crucial additional step also means that this mapping cannot be related to the usual free-to-harmonic canonical transformation in classical dynamics \cite{Goldstein}.

\added{Additionally, the transformation may have a significant impact on boundary conditions, depending on what is chosen. For example, if the free particle is embedded in a box of length $L$ with periodic boundaries, the condition $\phi(L,t)=\phi(0,t)$ transforms via Eq.~\eqref{eq:wavefunction} to produce 
\begin{equation}
 \psi\left(\frac{\cos(\omega\tau)}{\sqrt{\omega}}L,\tau\right)={\rm e}^{\frac{1}{4}i\sin(2\omega t)L^2}\psi(0,\tau),
\end{equation}
i.e. even the relatively simple case of periodic boundary conditions in the free case corresponds in the harmonic system to a non-trivial phase relation between the origin and a point in space that oscillates with time. It is interesting to observe that the boundary condition becomes tautological $\psi(0,\tau)=\psi(0,\tau)$, but only at the asymptotic times $t=\pm\infty$ in the original free particle picture.}

\section{Classical Case \label{sec:Classical}}
An important advantage of the preceding mapping is that it can be immediately replicated within the Koopman-von Neumann (KvN) framework. The KvN approach to classical dynamics is one which adopts the formal trappings of a Hilbert space theory \cite{PhysRev.40.749, Baker1958,Curtright2014,Groenewold1946,PhysRevA.92.042122}, employing four dynamical variables with the fundamental commutation relations:
	\begin{align}
	\left[\hat{x},\hat{\lambda}\right]=\left[\hat{p},\hat{\theta}\right]=i, \quad
	\left[\hat{\lambda},\hat{\theta}\right]=\left[\hat{\lambda},\hat{p}\right]=\left[\hat{\theta},\hat{x}\right]=0.
	\end{align}
The  Bopp operators \cite{AIHP_1956__15_2_81_0, doi:10.1086/288104, bondar_wigner_2013}  $\hat{\lambda}$ and $\hat{\theta}$  are introduced such that time evolutions are described by the Poisson bracket in phase space. \added{The generator of time translations is then the \emph{Koopman} (or equivalently the Liouvillian)  generating operator  $\hat{K}$}:
	\begin{align}
	\hat{K}=\hat{p}\hat{\lambda}-\hat{V}^{\prime}\left(\hat{x}\right)\hat{\theta} +f(\hat{x},\hat{p}), \label{eq:Koopmanoperator}
	\end{align}
where $f(\hat{x},\hat{p})$ is an arbitrary function of the phase space variables \cite{bondar_operational_2012}, which here we set to zero. This generator evolves the classical wavefunction $\left|\psi_{\rm cl}\right\rangle$ according to
	\begin{align}
	i\frac{{\rm d}}{{\rm d}t}\left|\psi_{\rm cl}\right\rangle =\hat{K}\left|\psi_{\rm cl}\right\rangle. 
	\end{align}

For a free particle, we have $\hat{K}=\hat{p}\hat{\lambda}$. Proceeding entirely in analogy with the quantum case, we first consider a squeezing procedure, using $\hat{U}_{\rm cl}(t)$:
\begin{align}
    \hat{U}_{\rm cl}(t)&={\rm e}^{i\frac{\ln(c(t))}{2}\left(\hat{p}\hat{\theta}+\hat{\theta}\hat{p}-\left(\hat{x}\hat{\lambda}+\hat{\lambda}\hat{x}\right)\right)}, \\
     \hat{U}_{\rm cl}^\dagger(t)\hat{x}\hat{U}_{\rm cl}(t)&=c(t)\hat{x}, \\ 
     \hat{U}_{\rm cl}^\dagger(t)\hat{p}\hat{U}_{\rm cl}(t)&=c^{-1}(t)\hat{p}, \\
      \hat{U}_{\rm cl}^\dagger(t)\hat{\lambda}\hat{U}_{\rm cl}(t)&=c^{-1}(t)\hat{\lambda},\\
       \hat{U}_{\rm cl}^\dagger(t)\hat{\theta}\hat{U}_{\rm cl}(t)&=c(t)\hat{\theta},\\
    -i \hat{U}_{\rm cl}^\dagger(t)\frac{\partial \hat{U}_{\rm cl}(t)}{\partial t}&=\frac{\dot{c}(t)}{2c(t)}\left(\hat{p}\hat{\theta}+\hat{\theta}\hat{p}-\left(\hat{x}\hat{\lambda}+\hat{\lambda}\hat{x}\right)\right).
\end{align}

Physically, for both free and harmonic oscillators this squeezing operator has a simple interpretation, namely that it scales the mass as $m\to c(t)m$. Such mass transformations are of some interest, owing to the fact that in a relativistic context, mass and proper time can be represented as conjugate variables. This means that the dynamics of the proper time can be represented in the Hamiltonian with a varying mass \cite{propertime}, which can itself be implemented via $\hat{U}_{\rm cl}$. The ability to construct a squeezing operator in a classical context also presents a number of opportunities to mimic quantum procedures. One example is the amplification of quantum Hamiltonians, which is achieved via squeezing \cite{Arenz2020amplificationof}, and could be replicated classically using $\hat{U}_{\rm cl}$.

If the squeezing operator is applied to $\hat{K}$, we obtain the transformed $\hat{K}^\prime$,
	\begin{align}
	\hat{K}^\prime=\frac{1}{c^2(t)}\left[\hat{p}\hat{\lambda}+\frac{c(t)\dot{c}(t)}{2}\left(\hat{p}\hat{\theta}+\hat{\theta}\hat{p}-\left(\hat{x}\hat{\lambda}+\hat{\lambda}\hat{x}\right)\right)\right]. \label{eq:Koopmanoperatorfirsttransform}
	\end{align}
\added{Application of squeezing to the Koopman operator has in this case resulted in  harmonic potential terms of the form $\hat{x}\hat{\lambda}+\hat{\lambda}\hat{x}$, but also additional $\hat{p}\hat{\theta}+\hat{\theta}\hat{p}$ terms . These in fact represent a minimal coupling to a vector potential, entirely in analogy with Eq.\eqref{eq:Htransformed}. While in the quantum case the minimal coupling is obtained via $\hat{p}\to \hat{p}-\hat{A}$, for KvN dynamics the coupling substitution is given by \cite{GOZZI2002152}:}
\begin{align}
    \hat{p}\hat{\lambda} \to  &\frac{1}{2}\left[\left(\hat{p}-\hat{A}\right)\hat{\lambda}+\hat{\lambda}\left(\hat{p}-\hat{A}\right)\right] \notag \\ +&\frac{1}{2}\frac{\partial \hat{A}}{\partial \hat{x}} \left[\left(\hat{p}-\hat{A}\right)\hat{\theta}+\hat{\theta} \left(\hat{p}-\hat{A}\right)\right],
\end{align}
\added{which for $\hat{A}=c(t)\dot{c}(t)\hat{x}$ yields $c^2(t)\hat{K}^\prime$. This suggests we should continue in an identical fashion as the previous quantum case, and apply a further transformation $\hat{R}_{\rm cl}$ designed to eliminate this vector potential}:
\begin{align}
   \hat{R}_{\rm cl}(t)&={\rm e}^{-ic(t)\dot{c}(t)\hat{x}\hat{\theta}}, \\
    \hat{R}_{\rm cl}^\dagger(t)\hat{p}\hat{R}_{\rm cl}(t)&=\hat{p}+c(t)\dot{c}(t)\hat{x}, \\
    \hat{R}_{\rm cl}^\dagger(t)\hat{\lambda}\hat{R}_{\rm cl}(t)&=\hat{\lambda}-c(t)\dot{c}(t)\hat{\theta}, \\
     i \hat{R}_{\rm cl}^\dagger(t)\frac{\partial \hat{R}_{\rm cl}(t)}{\partial t}& = (\dot{c}^2(t)+c(t)\ddot{c}(t))\hat{x}\hat{\theta}.
\end{align}
After some fairly tedious algebra, we obtain the sought for Koopman operator $\hat{K}^{\prime \prime}$:
\begin{equation}
    \hat{K}^{\prime \prime}=\frac{1}{c^2(t)}\left(\hat{p}\hat{\lambda}-\ddot{c}(t)c^3(t)\hat{x}\hat{\theta}\right).
\end{equation}
Once again, if one specifies $c(t)$ to have the form of Eq.\eqref{eq:scaletransform} and makes the time transformation given by Eq.\eqref{eq:ttransform} then $\hat{K}^{\prime\prime}$ becomes the generator for simple harmonic motion. Furthermore it is again possible to find a closed form for the combined unitary $\hat{\mathcal{U}}_{\rm cl}=\hat{U}_{\rm cl}(t)\hat{R}_{\rm cl}(t)$ due to the fact that 
\begin{equation}
    \left[\hat{p}\hat{\theta}+\hat{\theta}\hat{p}-\left(\hat{x}\hat{\lambda}+\hat{\lambda}\hat{x}\right),\hat{x}\hat{\theta}\right]=-4i\hat{x}\hat{\theta},
\end{equation}
meaning that Eq.\eqref{eq:BCH} can once again be applied to yield:
\begin{equation}
  \hat{\mathcal{U}}_{\rm cl}={\rm e}^{-i\ln(c(t))\left(c(t)\dot{c}(t)\left(1+\frac{c^2(t)+1}{c^2(t)-1}\right)\hat{x}\hat{\theta} +\frac{1}{2}\left(\hat{x}\hat{\lambda}+\hat{\lambda}\hat{x}-\hat{p}\hat{\theta}-\hat{\theta}\hat{p}\right)\right)}.
\end{equation}

One rather remarkable fact worth highlighting is that transformations of the form of $\hat{U}_{\rm cl}(t)\hat{R}_{\rm cl}(t)$ already exist in the KvN literature. Ref.~\cite{tdosc} demonstrates that this unitary (with suitably chosen $c(t)$) can be used to map a simple harmonic oscillator to one with a time dependent frequency. As has been made explicit here, this mapping can be extended even to the free particle provided an additional time coordinate transformation is performed. This of course raises the question of invariants, as it is easy to see that after the unitary transformations not only does the Koopman operator become explicitly time dependent, but so too does the Hamiltonian as the coordinate transformation is itself non-canonical. That is, while the transformation yields the same Koopman operator and dynamical equations of motion as the simple harmonic oscillator, it does \textit{not} have the same Hamiltonian, and that Hamiltonian is \textit{not} invariant. It is possible  however to show that in this case the time invariant operator $\hat{I}$ is given by \cite{tdosc,Sen2020}:
\begin{align}
    \hat{I}=&\frac{\hat{x}^2}{2c^2(t)}+\frac{1}{2}\left(\dot{c}(t)\hat{x} -c(t)\hat{p}\right)^2\notag \\ +&\frac{\hat{\theta}^2}{2c^2(t)} +\frac{1}{2}\left(\dot{c}(t)\hat{\theta}+c(t)\hat{\lambda}\right)^2.
\end{align}

\section{Mapping Open and hybrid System Evolutions \label{sec:open}}
Having demonstrated an almost identical procedure can be used to map a free particle to a harmonic oscillator in both quantum and classical dynamics, we now ask what effect this mapping will have when the evolution in question is no longer unitary, but has a dissipative component. In particular, we will consider a process which will cause decoherence in the position basis \cite{RevModPhys.75.715}. For quantum dynamics this is described by a Lindblad master equation~~\cite{Petruccione-open-systems-book}, given by
% \begin{align}\label{LindbladMasterEq}
% 	\frac{d\hat{\rho}(t)}{dt} &= -\frac{i}{\hbar} [\hat{H}, \hat{\rho}(t)] \notag\\
% 		&+ \frac{1}{2}\sum_{k=1}^K \left( 2 \hat{A}_k \hat{\rho}(t) \hat{A}_k^{\dagger} 
% 			-\left\{\hat{\rho}(t), \hat{A}_k^{\dagger} \hat{A}_k\right\} \right)
% \end{align}
\begin{align}\label{eq:LindbladMasterEq}
	\frac{d\hat{\rho}(t)}{dt} &= -\frac{i}{\hbar} [\hat{H}, \hat{\rho}(t)] 
		+ \frac{\gamma}{2}\left( 2 \hat{x} \hat{\rho}(t) \hat{x} 
			-\left\{\hat{\rho}(t), \hat{x}^2\right\} \right).
\end{align}
where $\hat{\rho}$ is the density matrix. The easiest way to interrogate the effect of the mapping on the additional dissipative terms is to exploit the fact that an infinitesimal evolution can be represented as a sum of two unitary maps \cite{EnsembleRank}:
\begin{align}\label{eq:mainformula}
    \hat{\rho}(t + {\rm d }t) =& \frac{1}{2} \left(
            {\rm e}^{-i\hat{H}_+{\rm d}t} \hat{\rho}(t) {\rm e}^{i\hat{H}_+{\rm d}t} 
            +
            {\rm e}^{-i\hat{H}_-{\rm d}t}  \hat{\rho}(t) \ {\rm e}^{-i\hat{H}_+{\rm d}t} 
        \right) \notag \\+& O\left({\rm d}t^2 \right),
\end{align}
where 
    \begin{equation}
    \hat{H}_\pm=\frac{1}{2}\hat{p}^2 \pm \sqrt{\frac{\gamma}{{\rm d}t}} \hat{x}.
\end{equation}
We can therefore assess the action of the mapping purely on $\hat{H}_\pm$ in order to ascertain its effect on the dissipative part of the action. Applying $\hat{\mathcal{U}}$, we obtain:
\begin{align}
    \hat{\mathcal{U}}^\dagger\hat{H}_\pm\hat{\mathcal{U}}&-i\hat{\mathcal{U}}^\dagger\frac{\partial \hat{\mathcal{U}}}{\partial t}= \notag \\ &\frac{1}{c^2(t)}\left(\frac{1}{2}\hat{p}^2+\frac{1}{2}\ddot{c}(t)c^3(t)\hat{x}^2\pm \sqrt{\frac{\gamma}{{\rm d}t}} c^3(t)\hat{x}\right).
\end{align}
If the time coordinate transformation ${\rm d}t = c^2(t){\rm d} \tau$ is also made, we find that the effect on the mapping can be expressed in terms of its effect on the dissipation parameter $\gamma \to c^4(\tau)\gamma$. Hence, a free system undergoing continuous measurement with strength $\gamma$ maps to a harmonic oscillator where the measurement has a time dependent strength $\cos^{-4}(\omega \tau)\gamma$.

It is relatively easy to show that the classical evolution that corresponds to the preceding quantum evolution is \cite{PhysRevA.92.042122}:
	\begin{align}
	i\frac{{\rm d}}{{\rm d}t}\left|\psi_{\rm cl}\right\rangle& =\hat{K}_{\rm FP}\left|\psi_{\rm cl}\right\rangle. \\
	\hat{K}_{\rm FP}&=\hat{\lambda}\hat{p}-iD\hat{\theta}^2
	\end{align}
which is itself nothing but a Fokker-Planck evolution with a constant dissipation parameter $D$. Application of $\hat{\mathcal{U}}_{\rm cl}$ yields 
\begin{align}
\hat{\mathcal{U}}_{\rm cl}^\dagger\hat{K}_{\rm FP}\hat{\mathcal{U}}_{\rm cl}&-i\hat{\mathcal{U}}_{\rm cl}^\dagger\frac{\partial \hat{\mathcal{U}}_{\rm cl}}{\partial t}= \notag \\ &\frac{1}{c^2(t)}\left(\hat{p}\hat{\lambda}-\ddot{c}(t)c^3(t)\hat{x}\hat{\theta}-ic^4(t)D\hat{\theta}^2\right).
\end{align}
Meaning that $D\to c^4(\tau)D$. Significantly, this means that the effect of the mapping on the dissipative part of the evolution is identical in both the classical and quantum scenarios, namely a scaling of the dissipation parameter by $c^4(\tau)$. 

\added{Finally, it is worth considering the free-harmonic mapping in the context of a quantum-classical hybrid system~\cite{Bondarquantumclassical}. A typical example would be a quantum system with non-commuting variables $\hat{z}=(\hat{x},\hat{p})$ coupled bilinearly to a classical one with commuting phase space classical variables $\hat{Z}=(\hat{X},\hat{P})$. An example hybrid Hamiltonian describing this scenario would be $\hat{H}_h$
\begin{equation}
    \hat{H}_h=\frac{1}{2}\hat{P}^2 +\frac{1}{2}\hat{p}^2 +k\hat{x}\hat{X}.
\end{equation}
The hybrid Liouvillian describing the evolution of this system is given by \cite{Bondarquantumclassical}} 
\begin{align}
    \hat{\mathcal{L}}_h &= \hat{H}_h+\frac{1}{2}\left(\frac{\partial \hat{H}_h}{\partial \hat{P}}\hat{\lambda}-\frac{\partial \hat{H}_h}{\partial \hat{X}}\hat{\theta} -\hat{Z}\cdot \nabla_Z\hat{H}_h\right) \\
    &=\hat{\lambda}\hat{P}+\frac{1}{2}\hat{p}^2-\frac{k}{2}\hat{x}\left(\hat{\theta}+\hat{X}\right),
\end{align}
\added{where $\hat{\lambda}$ and $\hat{\theta}$ are the Bopp operators associated with the classical variables $\hat{Z}$. Applying the appropriate unitary transformations to both the quantum and classical variables followed by the time coordinate transformation, we obtain a classical and quantum harmonic oscillator, together with the additional coupling term $\frac{k}{2}c^4(\tau)\hat{x}\left(\hat{\theta}+\hat{X}\right)$. As might be expected, the effect of the unitary mapping on the bilinear coupling is therefore the introduction of a time dependence into the coupling via $k\to c^4(\tau)k$.}
\section{Discussion \label{sec:Discussion}}
Taking as inspiration the existence of the coordinate mapping between the free particle and harmonic oscillator Schr{\"o}dinger equation, we have demonstrated that an equivalent transformation can be achieved in both quantum and classical dynamics via a combination of unitary operators and a time coordinate transform. This mapping can be further extended to time dependent frequency oscillators and open systems. The discontinuities in the time coordinate transformation also sheds some insight into how the continuous Hamiltonian (or Koopman) spectrum of a free particle can become the discrete oscillator spectrum. Moreover, it is also possible to map the free particle to an inverted oscillator with $\omega\to i\omega$. Significantly, an imaginary frequency \textit{removes} the discontinuities in the time transformation, and immediately implies that both the Hamiltonian and Koopman spectra for the inverted oscillator are continuous. 

Given the harmonic oscillator is perhaps \textit{the} paradigmatic model for any physical system, the existence of a novel mapping to it may be of some interest to a number of fields. A relevant example here would be that such a mapping would allow one to incorporate time dependent frequencies into both quantum and classical environment models, and the path integral methods used to solve them \cite{influencefunctional}. Additionally, the methods presented here can also be used to map a free particle to a linear potential, which can itself be mapped to a Lindblad-type model for entropic gravity \cite{PhysRevResearch.3.033065}. Connecting these two would therefore allow one to identify the unitary operators corresponding to the Kraus map for this model of entropic gravity.  Even divorced from application however, the existence of two almost identical mappings in both quantum and classical dynamics highlights the centrality of the harmonic oscillator model, and the unexpected commonalities it possesses between classical and quantum dynamics. 

\begin{acknowledgments}
This work was supported by Army Research Office (ARO) (grant W911NF-19-1-0377; program manager Dr.~James Joseph). The views and conclusions contained in this document are those of the authors and should not be interpreted as representing the official policies, either expressed or implied, of ARO or the U.S. Government. The U.S. Government is authorized to reproduce and distribute reprints for Government purposes notwithstanding any copyright notation herein.
\end{acknowledgments}

\end{document}